\documentclass[preprint,aps]{revtex4}
\usepackage{graphicx}
\usepackage{bm}
\usepackage{color}
\usepackage{subfigure}

\begin{document}
\baselineskip=15pt \parskip=5pt

\vspace*{3em}

\title{Colored Scalars And The CDF $W+$dijet Excess}

\author{ Tsedenbaljir Enkhbat$^1$, Xiao-Gang He$^{1,2,3}$, Yukihiro
Mimura$^1$, and Hiroshi Yokoya$^{1,4}$}

\affiliation
{$^1$Department of Physics and  Center for Theoretical Sciences, \\
National Taiwan University, Taipei, Taiwan 10617\\
$^2$INPAC, Department of Physics, Shanghai Jiao Tong University,
Shanghai, China 100871\\
$^3$National Center for Theoretical Sciences, National Tsing Hua University, Hsinchu 300,  Taiwan\\
$^4$National Center for Theoretical Sciences, National Taiwan
University, Taipei, Taiwan 10617}
\date{\today $\vphantom{\bigg|_{\bigg|}^|}$}

\begin{abstract}
The recent data on $W+$dijet excess reported by CDF may be interpreted
 as the associated production of a $W$ and a new particle of mass about
 150 GeV which subsequently decays into two hadron jets.
We study the possibility of explaining the $W+$dijet excess by colored scalar
 bosons.
There are several colored scalars which can have tree level
 renormalizable Yukawa couplings with two quarks, $({\bf 8}, {\bf 2},
 1/2)$, $(\bar{\bf 6}({\bf 3}), {\bf 3}({\bf 1}), -1/3)$, $(\bar {\bf
 6}({\bf 3}), {\bf 1}, -4/3(2/3))$.
If one of these scalars has a mass about 150 GeV, being colored it can
 naturally explain why the excess only shows up in the form of two
 hadron jets.
Although the required production cross section and mass put
 constraints on model parameters and rule out some possible scenarios
 when confronted with other existing data, in particular FCNC data, we
 find that there are strong constraints on the Yukawa couplings of these
 scalars.
 Without forcing the couplings to be some special texture forms
 most of the scalars, except the $({\bf 3}, {\bf 3}, -1/3)$, are in
 trouble with FCNC data.
We also study some features for search of these new particles at the RHIC
and the LHC and find that related information can help further to
 distinguish different models.
\end{abstract}

\maketitle

\section{Introduction}

The CDF collaboration has reported an excess in the production of two
jets in association with a $W$ boson production~\cite{CDF} from data
collected at the Tevatron with a center-of-mass energy of 1.96~TeV and
an integrated luminosity of 4.3~fb$^{-1}$.
The $W$ boson is identified through a charged lepton (electron or muon)
with large transverse momentum. The invariant mass of the dijet
system is found to be in the range of 120-160~{GeV}. The $W+$dijet
production has a few pb cross-section which is much larger than standard
model (SM) expectation.
The dijet system may be interpreted as an unidentified resonance with
mass around 150~{GeV} which predominantly decays into two hadron jets.
This leads to the speculation that a beyond SM new particle has
been found.
At present the deviation from the SM expectation is only at 3.2$\sigma$
level. The excess needs to be further confirmed. On the theoretical
side, our understanding of the parton distributions and related matter
still have room for improvement to make sure that the excess represents
genuine new physics beyond the SM~\cite{He:2011ss,Sullivan:2011hu}.
Nevertheless studies of new particle explanation has attracted much
attention.

Several hypothetic particles beyond SM have been proposed to explain the
CDF $W+$dijet excess, such as leptophobic $Z'$
model~\cite{leptoforbic,kingman}, technicolor~\cite{tech}, colored
vector, scalar~\cite{Wang:2011ta}, quasi-inert Higgs
bosons~\cite{Cao:2011yt} and the other possibilities~\cite{other}.
Common to all of these models is that the new particle must decay
predominantly into hadrons (dijet).
We note that a class of particles which can naturally have this
property.
These are those scalars which are colored and couple to quarks
directly.
In order for these scalars to be considered as a possible candidate, it
must satisfy constraints obtained from existing experimental
data.
Colored particles which couple to two quarks have been searched
for at the Tevatron and the LHC.
If the couplings to quarks/gluon are the same as the
QCD coupling, the color triplet diquark with a mass in the range
$290<m<630$~GeV is excluded at the Tevatron~\cite{Aaltonen:2008dn}, and
the mass intervals, $500 < m < 580$ GeV, $0.97 < m < 1.08$ TeV and $1.45
< m < 1.6$ TeV are excluded at the
LHC~\cite{:2010bc,Khachatryan:2010jd} whereas the LHC data is limited for
$m_{jj}>200$~GeV.
The color sextet diquarks with electric charge, $\pm 2/3$, $\pm 1/3$,
$\pm 4/3$, are excluded for their masses less than 1.8, 1.9, 2.7 TeV,
respectively~\cite{Han:2010rf}.
The color octet vectors/scalars which interact with quarks/gluon by QCD
coupling are excluded for $m<1.6$~TeV~\cite{Han:2010rf}.
If their couplings to
quarks/gluon are smaller than the QCD coupling the constraints are weaker.

Some aspects of colored scalars relevant to the CDF $W+$dijet data have
been considered recently~\cite{Wang:2011ta}.
In this work we carry out a systematic study to investigate the
possibility of colored scalar bosons $\eta$ as the new particle
explaining the CDF excess through $W\eta$ production followed by $\eta$
decays into two hadron jets.

At the tree level, there are several new scalar bosons which can have
renormalizable couplings to two quarks (or a quark and an anti-quark).
A complete list of beyond SM scalars which can couple to
SM fermions at the tree level~\cite{Davies:1990sc} and some of the
phenomenology have been studied
before~\cite{Davies:1990sc,DelNobile:2009st}.
The required production cross section and the mass from $W+$dijets
excess put constraints on model parameters. Some possible scenarios are
ruled out when confronted with other existing data, such as data from
flavor changing neutral current (FCNC) processes.
We find that without forcing of the Yukawa couplings to be some special
forms most of the scalars, except the $({\bf 3}, {\bf 3}, -1/3)$, are in
trouble with FCNC data.
We, however, do find that some other cases can be made consistent
with all data by tuning their couplings providing a possible explanation
for the $W+$dijet excess from CDF.
Justification of such choices may have a realization in a flavor model,
which is beyond the scope of the present work of phenomenology.
These colored scalars also have interesting signatures at the
Relativistic Heavy Ion Collider (RHIC) and the Large Hadron Collider
(LHC) which can be used to further distinguish different models.

The paper is organized as follows.
In Section~\ref{Sec:CS}, we study possible colored-scalars which can
couple to two quark (or a quark and a anti-quark), and determine their
Yukawa couplings by requiring that the colored scalar with a mass of 150
GeV to explain the CDF $W+$dijet excess data.
In Section~\ref{Sec:FCNC}, we study the constraints from FCNC data on
colored scalar couplings.
In Section~\ref{Sec:RHIC}, we give some implications for the RHIC and
the LHC.
Finally, we summarize our results in Sec.~\ref{Sec:Sum}.

\section{Colored Scalars and the CDF $W+$dijet Excess}\label{Sec:CS}

Scalar bosons which have color and have renormalizable Yukawa couplings
to two quarks or a pair of a quark and an anti-quark can be easily
determined by studying bi-products of two quarks~\cite{Davies:1990sc}.
The quarks transform under the SM $SU(3)_C\times SU(2)_L\times U(1)_Y$ as:
$q^i_L = (u^i_L, d^i_L)^T: ({\bf3}, {\bf2}, 1/6)$ the following bi-products,
$u^i_R: ({\bf3},{\bf1}, 2/3)$ and $d^i_R: ({\bf3},{\bf1}, -1/3)$.
Here $i$ and $j$ are generation indices.
With these quantum numbers, we can have the following quark bi-products
\begin{eqnarray}
&&\overline{u^i_R} q^j_L: ({\bf1} + {\bf8}, {\bf2}, - 1/2)\;,\;\;\;\;\;
\overline{ d^i_R} q^j_L: ({\bf1} + {\bf8}, {\bf2}, 1/2)\;,\nonumber\\
&&\overline{ q^{ic}_L} q^j_L: (\bar {\bf3} + {\bf6}, {\bf1} + {\bf3},
 1/3)\;,\;\;
\overline{ u^{ic}_R} u^j_R: (\bar {\bf3} + {\bf6}, {\bf1},
4/3)\;,\\
&&\overline{ u^{ic}_R} d^j_R: (\bar {\bf3} + {\bf6}, {\bf1},
 1/3)\;,\;\;\;\;\;\;\;\;
\overline{ d^{ic}_R} d^j_R: ( \bar {\bf3} + {\bf6}, {\bf1}, -2/3)\;,\nonumber
\end{eqnarray}
where the superscript ``$c$'' indicates charge conjugation.

For those scalars which only couple to right-handed quarks, the
contribution to $W$ associated production will be small because they do
not directly couple to $W$ boson.
To have large $W$ associated production for the CDF excess, we therefore
consider the following colored scalars which can couple to left-handed
quarks
\begin{eqnarray}
&&
\eta_8 = \sqrt{2}T^a\eta^a_8: ({\bf8},{\bf2},1/2), \nonumber\\
&&
\eta_{(6,3)} = K^a_{\alpha\beta}\eta^a_{(6,3)} : (\bar {\bf6}, {\bf3},
-1/3),
\qquad
\eta_{(6,1)} = K^a_{\alpha\beta}\eta^a_{(6,1)} : (\bar {\bf6}, {\bf1},
-1/3), \\
&&
\eta_{(3,3)} = \eta^\alpha_{(3,3)} : ({\bf3}, {\bf3}, -1/3),
\qquad \;\;\;\;\;\;
\eta_{(3,1)} = \eta^\alpha_{(3,1)} : ({\bf3}, {\bf1}, -1/3),\nonumber
\end{eqnarray}
where $\alpha$ is a color index, $T^a$ is the $SU(3)_C$ generator
normalized as ${\rm Tr}\,(T^aT^b)=\delta^{ab}/2$, and $K^a$
($a=1,\ldots,6$) is a generator of the symmetric tensor ($K^1_{11} =
K^2_{22} = K^3_{33}=1$, $K^4_{12} = K^4_{21} = K^5_{13} = K^5_{31} =
K^6_{23}=K^6_{32}=1/\sqrt2$).
The color component fields, $\eta^a_8$, $\eta^a_{(6,3)}$ and
$\eta^a_{(6,1)}$, are defined by having the kinetic energy term
normalized properly.

We denote the component fields of $SU(2)_L$ as follows:
\begin{eqnarray}
\eta_8^A = \left (
		\begin{array}{c}
			\eta_8^0\\
			\eta_8^-
		\end{array}
	\right ), \
\eta_{(6,3)}^A{}_B = \left (
		 \begin{array}{cc}
		  \eta_{(6,3)}^{-1/3}/\sqrt{2}  & \eta_{(6,3)}^{2/3}\\
		  \eta_{(6,3)}^{-4/3}  &-\eta_{(6,3)}^{-1/3}/\sqrt{2}
		 \end{array}
	      \right ), \
\eta_{(3,3)}^A{}_B = \left (
		 \begin{array}{cc}
		  \eta_{(3,3)}^{-1/3}/\sqrt{2}  & \eta_{(3,3)}^{2/3}\\
		  \eta_{(3,3)}^{-4/3}  &-\eta_{(3,3)}^{-1/3}/\sqrt{2}
		 \end{array}
	      \right ),
\end{eqnarray}
where $A,B$ are the $SU(2)_L$ indices.
For neutral $\eta^0$, the physics component can be separated according to
their parity property with $\eta^R = \sqrt{2}\, {\rm Re}\,(\eta^0)$
and $\eta^I = \sqrt{2}\, {\rm Im}\,(\eta^0)$.

\begin{figure}[t]
\begin{center}
\includegraphics[width=.5\textwidth,clip]{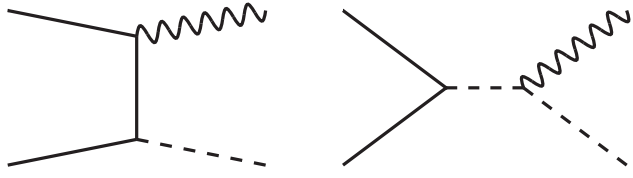}
\caption{Feynman diagrams for the $t$-channel and $s$-channel process in
 $W+\eta$ production.
Solid, wavy and dashed line represents quark or anti-quark, $W^{\pm}$
 and $\eta$, respectively.
\label{Feyn}}
\end{center}
\end{figure}

For $W\eta$ production by $p \bar p$ collision, the leading
contributions are from the $t$-channel and $s$-channel tree diagrams as
shown in Fig.~\ref{Feyn}.
For $\eta_8$, $\eta_{(6,3)}$ and $\eta_{(3,3)}$ the $s$-channel
production can exist in addition to the diagram of $t$-channel quark
exchange.
One needs to know how the colored scalars couple to quarks and the $W$
boson.
We list the Yukawa couplings in the quark mass eigenstate basis in the
following,
\begin{eqnarray}
-L_{\eta}
&=&
\overline{U_R} Y_{8u} \eta^a_8{}_A T^a Q_L^A
+
\overline{Q_L}_A Y_{8d} \eta^{aA}_8 T^a  D_R  \nonumber
\\
&+&
\frac12 \overline{Q_L^c}_A Y_{(6,3)} \eta_{(6,3)}^{aA}{}_B  K^a Q_L^B
+
\frac12 \overline{Q_L^c}_A Y_{(6,1)} \eta_{(6,1)}^a K^a Q_L^A
\\
&+&
\frac12 \overline{Q_L^c}_A^\alpha Y_{(3,3)} \eta_{(3,3)}^{\beta A}{}_B
Q_L^{\gamma B} \epsilon_{\alpha\beta\gamma}
+
\frac12 \overline{Q_L^c}^\alpha_A Y_{(3,1)} \eta_{(3,1)}^\beta
Q_L^{\gamma A} \epsilon_{\alpha\beta\gamma}
+ h.c. \nonumber
\end{eqnarray}
where $\eta_A = (i\sigma_2)_{AB} \eta^B$.
The flavor space is described as
$U_R = (u^i_R)^T$, $D_R = (d^i_R)^T$ and
$Q_L = (q^i_L)^T = ( u^i_L, V_{ij} d^j_L)^T$,
%
%
where $V$ is a CKM quark mixing matrix,
and
$Y^{ij}_I$ ($I=$\,{8$q$, (6,3), (6,1), (3,3), (3,1)}) are the coupling
matrix in flavor space.
$Y_{(6,3)}^{ij}$ and $Y_{(3,1)}^{ij}$ are symmetric,
and,
$Y_{(6,1)}^{ij}$ and $Y_{(3,3)}^{ij}$ are anti-symmetric,
under the exchange of flavor indices $i$ and $j$.
The diquark couplings are
\begin{eqnarray}
-L_{\eta_{\rm tri}}
&=&
\frac12 \overline{Q_L^c} Y_{\rm tri} \eta_{\rm tri} Q_L + h.c. \nonumber\\
&=& \frac12 \overline{U_L^{c}} Y_{\rm tri} \eta_{\rm tri}^{-4/3} U_L
  - \frac1{\sqrt2} \overline{U_L^{c}} Y_{\rm tri} V \eta_{\rm tri}^{-1/3} D_L
      + \frac12 \overline{D_L^{c}} V^T
             Y_{\rm tri} V \eta_{\rm tri}^{2/3} D_L +h.c. \,, \label{di-int}\\
 -L_{\eta_{\rm s}}
&=&
\frac12 \overline{Q_L^c} Y_{\rm s} \eta_{\rm s} Q_L + h.c.
=
  \overline{U_L^{c}} Y_{\rm s} V \eta_{\rm s}^{-1/3} D_L +h.c. \,,\nonumber
  \label{di-int-s}
\end{eqnarray}
where
$\eta^Q_{\rm tri}$ is $SU(2)_L$ triplet,
and
$\eta^Q_{\rm s}$ is $SU(2)_L$ singlet.

The electroweak gauge interactions are given by
\begin{eqnarray}
L_W &=& i\overline{Q_L} \gamma_\mu {\cal D}^\mu Q_L
+
i\overline{U_R} \gamma_\mu {\cal D}^\mu U_R
+
i\overline{D_R} \gamma_\mu {\cal D}^\mu D_R
+
({\cal D}^\mu \eta_I)^\dagger ({\cal D}_\mu \eta_I)\;,
\end{eqnarray}
where ${\cal D}_\mu$ is the covariant derivative.
The electroweak gauge interactions of the colored scalars are obtained
from the following:
\begin{eqnarray}
({\cal D}^\mu \eta_8)^\dagger ({\cal D}_\mu\eta_8)
&=&
\left|\left(i\partial_\mu + \left({1\over 2} - s^2_W\right) g_Z Z_\mu +
       e A_\mu\right)\eta^+_8 +
{1\over \sqrt{2}}g W^+_\mu \eta^0_8\right|^2 \nonumber\\
&+&
\left|\left(i\partial_\mu -{1\over 2} g_Z Z_\mu\right)\eta^0 +
{1\over \sqrt{2}} g W^+_\mu \eta^-\right|^2\;,\nonumber
\\
({\cal D}^\mu \eta_{\rm tri})^\dagger ({\cal D}_\mu \eta_{\rm tri})
&=&
\left|\left(i\partial_\mu + \left(1-{2\over 3} s^2_W\right)g_Z Z_\mu +
{2\over 3} e A_\mu\right)\eta_{\rm tri}^{2/3} +
g W^+_\mu \eta^{-1/3}_{\rm tri}\right|^2\nonumber
\\
&+&
\left|\left(i\partial_\mu + {1\over 3} s^2_W g_Z Z_\mu -
{1\over 3} e A_\mu\right)\eta^{-1/3}_{\rm tri}  +
gW^-_\mu \eta^{2/3}_{\rm tri} +
g W^+_\mu \eta^{-4/3}_{\rm tri}\right|^2\nonumber
\\
&+&
\left|\left(i\partial_\mu \left(-1+{4\over 3}s^2_W\right)g_Z Z_\mu
- {4\over 3} e A_\mu\right)\eta^{-4/3}_{\rm tri}
+ g W^-_\mu \eta^{-1/3}_{\rm tri}\right|^2\;,
\\
({\cal D}^\mu \eta_{\rm s})^\dagger ({\cal D}_\mu \eta_{\rm s})
&=&
\left|\left(i\partial_\mu + {1\over 3} s^2_W g_Z Z_\mu -
{1\over 3} e A_\mu\right)\eta_{\rm s}^{-1/3} \right|^2\;,
\nonumber
\end{eqnarray}
where $s_W$ is the sine of the Weinberg angle $\theta_W$, and $g_Z = e/s_W
c_W$.
In the above color indices are suppressed and interaction with gluons
are omitted.

For $\eta_8$ color-octet and $\eta_{(6,3)}$, $\eta_{(3,1)}$ diquarks,
the dominant contributions to the $W\eta$ production
are from $\eta$ couplings to the first generation, $Y^{11}_I$.
For $\eta_{(6,1)}$, $\eta_{(3,3)}$ diquarks, on the other hand,
because the Yukawa coupling matrix is anti-symmetric in generation space,
the dominant contribution would come from $Y^{12}_I$ term
(which includes $u$ and $d$ quark coupling suppressed by Cabibbo mixing).

In general different component in $\eta$ can have different masses.
In order to avoid the contribution to $\rho$ parameter, we assume that
all the components have the same masses for simplicity.

\begin{figure}[t]
\begin{center}
\includegraphics[width=.5\textwidth,clip]{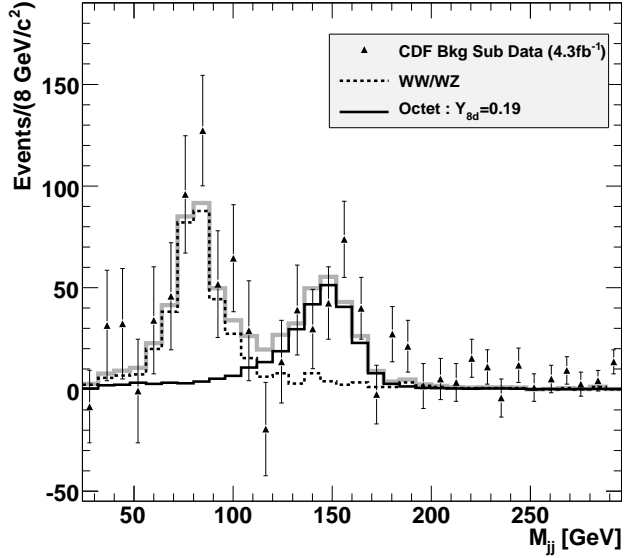}
\caption{Dijet invariant-mass distribution in $W+$dijets events at the
 Tevatron.
CDF data taken from Ref.~\cite{CDF} is shown with our MC simulation
 results for the color-octet scalar model with $Y_{8d}=0.19$ (solid)
 and the standard-model $WW+WZ$ contribution (dashed).
\label{CDF}}
\end{center}
\end{figure}

Since interactions and the masses of the colored scalars are fixed, the
only unknowns parameters, the Yukawa couplings, can be determined by
requiring the colored scalars to explain the CDF $W+$dijet data.
We consider the different type of colored scalars separately.
For the fit, we use {\tt
MadGraph/MadEvent}~\cite{madgraph,Alwall:2007st} and {\tt
Pythia}~\cite{pythia} for the particle-level event-generation, and {\tt
PGS} for the fast detector simulation.
Jets are defined in cone algorithm with $R=0.4$.
We apply the same kinematical cuts as those denoted in Ref.~\cite{CDF}.
The reconstructed jet momenta are rescaled so that the dijet
invariant-mass has correct peak at the resonance masses.
The simulation result for the case of color-octet scalars $\eta_8$ with
$Y_{8d}$ coupling is shown in Fig.~\ref{CDF}.
Inclusive $W+\eta$ production cross-section at the Tevatron is estimated
to be 2~pb (without multiplying $K$-factor).
For other cases, we obtained similar distribution.
We list the central values of Yukawa couplings for each case in the
following,
\begin{eqnarray}
 &&Y^{11}_{8u} = 0.13\;,\;\;\;\;\;\;Y^{11}_{8d} = 0.19\;,\nonumber \\
 &&Y^{11}_{({6},3)} = 0.32\;,\;\;\;\;Y^{11}_{(3,1)} = 0.5\;,\nonumber \\
 &&Y^{12}_{({6},1)} = 1.0\;,\;\;\;\;\;\;\;\;\;Y^{12}_{(3,3)} = 0.62\,.
  \label{central-value}
\end{eqnarray}
We see that the Yukawa couplings are of order ${\cal O}(10^{-1})$ for
color-octet scalars, but close to ${\cal O}(1)$ for diquark scalars.
The sizeable Yukawa couplings for the diquark scalars come from the
fact that the Tevatron is a $p\bar{p}$ collider therefore the production
of diquark must pick up one sea-quark whose distribution function is
suppressed.
The large values for $Y_{(6,1)}^{12}$ and $Y_{(3,3)}^{12}$ couplings are
required since the production cross-section is suppressed due to the
Cabibbo mixing suppressed couplings to $u$ and $d$ quarks and the
suppressed $s$ or $c$-quark parton distribution inside a proton.
The difference between them is mainly due to being a triplet or a
singlet under $SU(2)_L$.

Two comments are in order about the sizeable colored scalar Yukawa couplings
which may cause problems in decay widths and constraints from direct resonant search for these scalers, at experiments such as at the UA2~\cite{UA2}.
First, the decay widths of these scalars are less than 1~GeV for
color-octet cases, and a few to several GeV for diquark scalars, where
the flavour structure of the Yukawa couplings of the scalars to quarks
is assumed to those determined in the next section.
These decay widths are small enough to regard the width of the
observed dijet resonance as the consequence of the resolution of the jet
momentum measurements.
Second, inclusive production of the scalars which couple to quarks are
constrained by the two-jet invariant mass spectrum measurement in the
UA2 experiment~\cite{UA2}.
For $m_{jj}\simeq150$~GeV, the cross-section times the branching ratio
to two jets is excluded for $\sigma\cdot{\mathcal B}\gtrsim 80$~pb.
The couplings in Eq.(8) provide values for $\sigma\cdot{\mathcal B}$ in pb as
\begin{equation}
 \eta_8 : 41 \ ({\rm for~}Y_{8u}^{11}),\ \ 34 \ ({\rm for~}Y_{8d}^{11}), \ \
  \eta_{(6,3)}: 14,\ \ \eta_{(6,1)}:  78,\ \
  \eta_{(3,1)}: 29,\ \  \eta_{(3,3)}: 33 \ .
\end{equation}
%
%
From the above values we see that  the couplings in Eq.(8) cannot be excluded by
the UA2 measurement. The cross section for $\eta_{(6,1)}$ is on the border
of the constraint.

We can estimate the $Z+$dijet production cross-section at the Tevatron.
For the couplings in Eq.~(\ref{central-value}), the $Z$+$\eta$ production
cross-sections are estimated to be
\begin{equation}
 \eta_8 : 0.16 \ (Y_{8u}^{11}),\ \ 0.25 \ (Y_{8d}^{11}), \ \
  \eta_{(6,3)}: 0.14,\ \ \eta_{(6,1)}:  0.69,\ \
  \eta_{(3,1)}: 0.67,\ \  \eta_{(3,3)}: 0.38 \ [{\rm pb}].
\end{equation}
The largest $\sigma(Z\eta)$ is about 0.7 pb,
which is 23\% of $\sigma(ZZ+ZW)$ within the SM estimation in
leading-order.
This fraction is similar to $\sigma(W\eta)/\sigma(WW+WZ)\sim
(2\;\mbox{to}\;4)/18$.
Therefore, although there have been no statistically significant signal
on the diboson production in $\ell^{-}\ell^{+}jj$ mode at the Tevatron
yet, $Z+\eta$ production should be more carefully studied.

\section{FCNC Constraints On Colored Scalars}\label{Sec:FCNC}

From the previous section we see that the Yukawa couplings of these
colored scalars to the first and second generations are much
larger than that of the usual Higgs in order to explain the CDF $W+$dijets excess. Therefore
we need to check whether such large Yukawa couplings are consistent with data. We now study
constraints from new FCNC interactions by colored scalars which may induce sizeable
meson-antimeson mixing. We consider each case separately in the following.

\subsection{Octet-doublet scalar}

Some phenomenological studies of the octet-doublet scalar can be found in
\cite{Davies:1990sc,DelNobile:2009st,Manohar:2006ga}.
Here we study the constraints from the mixing of mesons for
large Yukawa coupling to the first generation of quarks.

For $\eta_8$ couples with $U_R$ and $Q_L$, we have
\begin{eqnarray}
\overline{U_R} Y_{8u} \eta_8{}_A Q_L^A
=
\bar U_R Y_{8u} \eta^0_8 U_L  - \bar U_R Y_{8u} V \eta^+_8 D_L .
\end{eqnarray}
If $Y_{8u}$ is not diagonal,
exchange of $\eta^0_8$ will induce large FCNC effects at tree level,
such as $D^0$-$\bar D^0$ mixing, making the model inconsistent.
Even if $Y_{8u}$ is diagonal,
exchange of $\eta^+_8$ at loop level
can also induce FCNC interaction which may result in
too large $K^0$-$\bar K^0$ and $B^0_{d,s}$-$\bar B^0_{d,s}$ mixings.
To minimize possible FCNC interaction, we will work with a special case
where $Y_{8u} = y_{8u} I$ (where $I$ is a unit matrix) for illustration
($Y_{8u}^{ij}=y_{8u}\delta^{ij}$).
\begin{figure}[t]
\begin{center}
\includegraphics[width=.75\textwidth,clip]{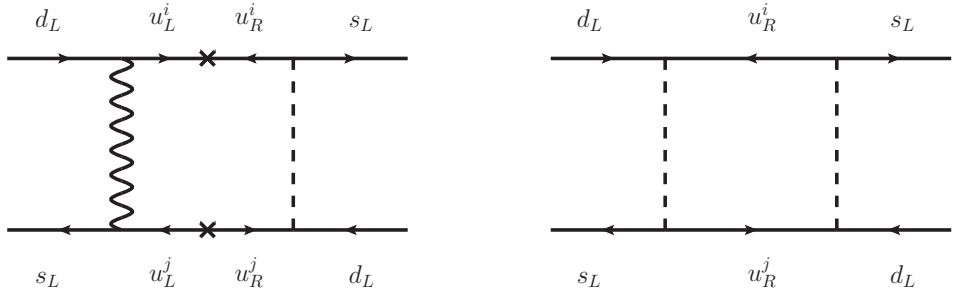}
\caption{Box diagrams for the $K^0$-$\bar K^0$ mixing
induced by $Y_{8u}$ coupling.
Dashed line represents the octet-doublet scalar propagation.
\label{FCNC}}
\end{center}
\end{figure}
In the case where only $Y_{8u}$ coupling is turned on
($Y_{8d}=0$),
the $K^0$-$\bar K^0$ mixing operator
$(\overline{d_L} \gamma_\mu s_L)(\overline{d_L} \gamma^\mu s_L)$
is induced by $W$-$\eta$ and $\eta$-$\eta$ box diagrams
shown in Fig.~\ref{FCNC}.

In order to show the $\eta$ contribution,
we define the following quantities:
\begin{eqnarray}
E_1 &=& \sum_{i,j} \lambda_i^s \lambda_j^s
z_W \sqrt{z_i z_j}\, G(z_i,z_j,z_W), \nonumber\\
E_2 &=& \sum_{i,j} \lambda_i^s \lambda_j^s
z_W
F(z_i,z_j,z_W),\nonumber \\
E_3 &=& \sum_{i,j} \lambda_i^s \lambda_j^s
\sqrt{z_i z_j}\,
F(z_i,z_j,z_W), \\
E_4 &=& \sum_{i,j} \lambda_i^s \lambda_j^s
z_i z_j\,
G(z_i,z_j,z_W),\nonumber \\
E_5 &=& \sum_{i,j} \lambda_i^s \lambda_j^s
z_W
F(z_i,z_j,1),\nonumber
\end{eqnarray}
where $z_i = m_{u_i}^2/m_\eta^2$,
$z_W = M_W^2/m_\eta^2$
($m_\eta$ is the mass of $\eta$ boson),
and $\lambda^s_i=V^{*}_{id}V_{is}$.
The loop functions $F$ and $G$ are given by
\begin{eqnarray}
&&
F(x,y,z) =
\frac{x^2\log(x)}{(1-x)(x-y)(x-z)}+
\frac{y^2\log(y)}{(1-y)(y-x)(y-z)}+
\frac{z^2\log(z)}{(1-z)(z-y)(z-x)},\nonumber\\
&&
G(x,y,z) =
\frac{x\log(x)}{(1-x)(x-y)(x-z)}+
\frac{y\log(y)}{(1-y)(y-x)(y-z)}+
\frac{z\log(z)}{(1-z)(z-y)(z-x)}.
\end{eqnarray}
We
obtain the $\eta$ contribution of $K^0$-$\bar K^0$ mixing amplitude as
\begin{eqnarray}
M_{12}^{K}(\eta)
& =&
 \frac{f_K^2m_K \hat{B}_1}{48 \pi^2 M^2_W}
 \left(
   \frac13 \frac{y_{8u}^2g_2^2}{2} \left(-E_1 + \frac18 E_3 \right)
   + \frac{11}{18} \frac{y_{8u}^4}{8} E_5
\right),
\end{eqnarray}
where $f_K$ is a kaon decay constant,
 $m_K$ is a kaon mass,
and
 $\hat{B}_1$ is a bag parameter
 from the matrix element of the
 $(\overline{d_L} \gamma_\mu s_L)(\overline{d_L} \gamma^\mu s_L)$
 operator between $K$ mesons~\cite{Laiho:2009eu}.
Here the $E_1$ and $E_3$ terms are from the $W$-$\eta$ box diagram
while the $E_5$ term is from the $\eta$-$\eta$ box diagram.

The kaon mass difference is obtained by $\Delta m_K = 2 |M_{12}^K({\rm
full})|$.
Inserting the value $Y^{11}_{8u}$ given in Eq.~(\ref{central-value})
(under the current assumption, $y_{8u} = Y^{11}_{8u}$),
we find that the $E_1$ term gives dominant contribution,
which corresponds to the $W$-$\eta$ box diagram with charm quark
mass insertions.
The short distance SM contribution has uncertainty
which mainly comes from the charm mass and QCD correction.
For the $K^0$-$\bar K^0$ system,
the short distance SM contribution with the next-to-leading order QCD correction
can fill roughly 80\%
of the experimental result, $\Delta M_K= 3.483 \times 10^{-15}$ GeV~\cite{pdg}.
Though the long distance contribution is hard to be estimated,
the total mixing amplitude in SM can be consistent with the experiment.
We exhibit the ratio of the leading order $\eta$ contribution
and the short distance SM contribution,
which is free from the hadronic uncertainty,
\begin{eqnarray}
&&
\frac{M_{12}^K(\eta)}{M_{12}^K({\rm SM})} \simeq
0.12\times
\left(\frac{y_{8u}}{0.13}\right)^2
\left(\frac{150\,\mbox{GeV}}{m_\eta}\right)^4.
\end{eqnarray}
Here we see that  the contributions from the octet scalar
is at 12\% of the short distance SM contribution of the mixing amplitude
for the value $Y_{8u}^{11}  \simeq0.13$
suggested by the $W+$dijet excess,
and therefore, consistent with the experimental result of kaon mass difference.
The imaginary part of the mixing amplitude gives indirect CP violation
in $K^0$-$\bar K^0$ mixing.
We find that ${\rm Im}\, M_{12}^K(\eta)/{\rm Im}\,M_{12}^K({\rm SM})$
is less than 2\%, and thus it is consistent with experiments.
The mixing amplitudes of
$B_{s,d}^0$-$\bar B_{s, d}^0$
are
obtained just by replacing $\lambda_i^s$, $f_K$, $m_K$ and $\hat{B}_1$ properly,
and they are
found to be
at the level of less than 1\%
of the standard model prediction as well as the experimental result.

When $\eta_8$ couples to $D_R$ and $Q_L$, we have
\begin{eqnarray}
\overline{Q_L}_A Y_{8d} \eta^{A}_8  D_R
= \overline{U_L} Y_{8d} \eta_8^+ D_R + \overline{D_L} V^\dagger Y_{8d} \eta^0_8 D_R .
\end{eqnarray}
In this case, to avoid large tree level FCNC,
one is forced to have $V^\dagger Y_{8d}$ to be diagonal.
Also similar to the previous case to avoid potential
large one loop FCNC, we make our illustration,
of the form ($Q \to Q^\prime = V^\dagger Q$)
\begin{eqnarray}
\overline{Q_L^\prime}_A Y_{8d} \eta_8^A D_R
 = \overline{U_L} V Y_{8d} \eta^+_8 D_R  + \overline{D_L} Y_{8d} \eta^{0}_8 D_R ,
\end{eqnarray}
with $Y_{8d} = y_{8d} I$.
In this case,
the $\eta$ contribution is
\begin{eqnarray}
M_{12}^{K}(\eta)
& =&
 \frac{f_K^2m_K }{48 \pi^2 M^2_W}
 \left(
   \frac{-\hat{B}_4+\hat{B}_5}8 \left(\frac{m_K}{m_s+m_d} \right)^2
   \frac{y_{8d}^2g_2^2}{2} \left(-E_2+\frac12 E_4\right)
  + \frac{11}{18} \hat{B}_1\frac{y_{8d}^4}{8} E_5
 \right),
\end{eqnarray}
where $\hat{B}_4$ and
$\hat{B}_5$ operators are
the bag parameters from the matrix elements of the operators,
 $(\overline{d_R}  s_L)(\overline{d_L} s_R)$
and
 $(\overline{d_R}^\alpha  s_{\beta L})(\overline{d_L}^\beta s_{\alpha R})$
between $K$ mesons, respectively~\cite{Gupta:1996yt}.
The contribution is small ($O(1)$\% of the experimental value) for the
 value for the Yukawa coupling chosen from $W+$dijet excess.
The situation is the same for the $B^0$-$\bar B^0$ mixing amplitude.

The $D^0$-$\bar D^0$ mixing amplitude is obtained
by exchanging $y_{8u} \leftrightarrow y_{8d}$,
and replacing $m_{u_i} \to m_{d_i}$
and $\lambda^s_i \to V_{ui} V_{ci}^*$
in the expressions of $K^0$-$\bar K^0$ mixing.
The mixing amplitude of
$D^0$-$\bar D^0$
induced by the $Y_{8u} = y_{8u}I$ coupling
is found to be very small at the level of less than $10^{-3}$
compared to the short distance SM contribution.
On the other hand, $D^0$-$\bar D^0$ mixing amplitude
induced by the $Y_{8d}=y_{8d}I$ coupling
receives a
large $W$-$\eta$ box contribution (corresponding to the $E_1$ term),
which is comparable to the short distance SM one.
However,
the short distance SM contribution of $D^0$-$\bar D^0$ mixing is tiny
compared to the experimental result,
\begin{equation}
\frac{2|M_{12}^D (\eta)|}{\Delta m_D^{\rm exp}}
\simeq5.5\times 10^{-4}
\left(\frac{y_{8d}}{0.19}\right)^2
\left(\frac{150\,\mbox{GeV}}{m_\eta}\right)^4.
\end{equation}
It is expected that long distance contribution in the SM can produce the experimental value.
For our purpose, it is therefore safe to say that $Y_{8d}$ coupling required by
the CDF $W+$dijet data can satisfy constraints from $D^0$-$\bar D^0$ mixing data

Note that the octet-doublet scalar with the form $Y^{ij}_{8d} = y_{8d}\delta^{ij}$ can
decay into $b\bar{b}$, giving 20\% of $b$-jet pair fraction in $W+\eta$
events.

One can also try to keep both $Y_{8u}$ and $Y_{8d}$ simultaneously non-zero.
But there is a large contribution to $b_L \to s_R \gamma$ ($b_R \to s_L \gamma$)
amplitude proportional to
$Y_{8u}^{33}Y^{22}_{8d} V_{ts}$ ($Y_{8u}^{33}Y^{33}_{8d} V_{ts}$).
This combination must be small resulting
in one of the $|Y^{11}_{8u,8d}|$ to be much smaller than the other.
This virtually goes back to the previous two cases studied.

One may be able to forbid
one of the $Y^{8u,8d}$ couplings by some discrete $Z_2$ symmetry, such as $\eta_8 \ to -\eta_8$ and $U_R \to - U_R$ and $D_R \to D_R$
to eliminate $Y^{8d}$. But to have $Y^{8u}$ proportional to unit chosen earlier, this raises a question how natural the choice is.
While this may be achievable by some flavor symmetry to enforce the special texture form,
such endeavor is beyond scope of this work and we will confine ourselves to phenomenological study only.  We conclude that the cases with $\eta_8$ couples to either
$U_R$ only or $D_R$ only is a phenomenologically viable model.

\subsection{Color sextet and triplet diquarks}

Now let us study if the color sextet or triplet diquarks are allowed.
Some phenomenological studies of the color sextet and triplet scalars can be found in
\cite{Davies:1990sc,DelNobile:2009st,Ma:1998pi,Mohapatra:2007af,Chen:2008hh,Ajaib:2009fq}.

\subsubsection{The $({\bf 6},{\bf 3},1/3)$ scalar}

The sextet diquark $\eta_{(6,3)}$ with Yukawa couplings required to
explain CDF $W+$dijet excess will lead to too large mixing in
$D^0$-$\bar D^0$ and $K^0$-$\bar K^0$ in contradiction with data. From Eq.~(\ref{di-int})
one can see that exchange of
$\eta_{(6,3)}^{-4/3}$ at tree level can generate a mixing amplitude
for $D^0$-$\bar D^0$ if $Y_{(6,3)}^{11}Y_{(6,3)}^{22}\neq 0$.
The constraint is estimated
as $Y_{(6,3)}^{11}Y_{(6,3)}^{22} \alt 10^{-7} (m_\eta^2/(150\, {\rm
GeV}))^2$.
Tree level mixing for $K^0 - \bar K^0$ is also generated by
$\eta_{(6,3)}^{2/3}$ exchange.
These mixing contributions can be eliminated by letting
$Y^{22}_{(6,3)}=(V^T Y_{(6,3)} V)_{22} = 0$
by choosing $Y^{12}_{(6,3)}= Y^{11}_{(6,3)} \tan\theta_C/2$,
where $\theta_C$ is a Cabibbo mixing angle.
However, under the exact cancellation of the tree level contributions,
the loop level contributions are too large.
One then has to arrange cancellation between the tree and one loop contributions. This may represents a problem of fine tuning.
Although this appears quite unnatural and harder to realize for building a model
compared to the octet case, from purely phenomenological point of view it is not ruled out yet.

The other diquarks $\eta_{(6,1),(3,3),(3,1)}$
do not induce the tree-level meson-antimeson mixing,
but can be generated at the 1-loop level through the box diagram.

\subsubsection{The $({\bf6},{\bf1},1/3)$ scalar}

Because the diquark $\eta_{(6,1)}$ is an $SU(2)_L$ singlet
and
$Y_{(6,1)}^{ij}$ is anti-symmetric. The CDF $W+$dijet excess requires a large value of
$Y_{(6,1)}^{12}$.
For illustration, let us consider a simple case with
$Y^{12}_{(6,1)}\neq0$ and
$Y_{(6,1)}^{23} = Y_{(6,1)}^{13}=0$ in the $Q^\prime = (V^\dagger U_L,D_L)$ basis:
\begin{eqnarray}
-L = \frac12 \overline{Q^\prime_L{}^c} Y_{(6,1)} \eta_{(6,1)}^a K^a Q^\prime_L
=
Y^{12}_{(6,1)}
(V_{i1}^* \overline{u^i_L{}^c}  \eta_{(6,1)}^a K^a  s_L -
   V_{i2}^* \overline{u^i_L{}^c}  \eta_{(6,1)}^a K^a d_L).
\end{eqnarray}
The contribution to $K^0$-$\bar K^0$ mixing amplitude from $\eta_{(6,1)}$ is
\begin{eqnarray}
M_{12}^{K}(\eta)
 =
 \frac{f_K^2m_K }{48 \pi^2 M^2_W}
 \left(  \frac{(Y^{12}_{(6,1)})^2 g_2^2}{4}\hat{B}_1
           \left(- E_2 +\frac12 E_4  \right)
 + \frac{15 \hat{B}_4 + \hat{B}_5}{16}
 \left(\frac{m_K}{m_s+m_d} \right)^2
 \frac{(Y^{12}_{(6,1)})^4}8 E_5
\right).
\end{eqnarray}
For $Y_{(6,1)}^{12}=1$ and $m_\eta=150$ GeV as required by the CDF $W+$dijet data,
it gives twice that of the short distance SM contribution constructively,
due to the enhancement factor $(m_K/(m_s+m_d))^2$.
The coupling $Y_{(6,3)}^{12}$ can also cause an excess of strangeless charm decay, such as
$D\to \pi\pi$.
The interaction generates a strangeless charm decay operator,
\begin{equation}
-\frac{(Y^{12}_{(6,1)})^2 V_{cd} V_{ud}^*}{4m_\eta^2}
\left[
(\overline{u_L}^\alpha \gamma_\mu c_{\alpha L})(\overline{d_L}^\beta \gamma^\mu d_{\beta L})
+
(\overline{u_L}^\alpha \gamma_\mu c_{\beta L})(\overline{d_L}^\beta \gamma^\mu d_{\alpha L})
\right],
\end{equation}
where we use
\begin{equation}
(K^a)_{\alpha\beta}(K^a)^{\delta\gamma} =
\frac12 (\delta_\alpha^\gamma\delta_\beta^\delta+\delta_\alpha^\delta\delta_\beta^\gamma).
\end{equation}

The contribution interfere with the standard model amplitude
at 40\% (including the color suppressed process) for $Y^{12}_{(6,3)}= 1$
and $m_\eta = 150$ GeV,
which contradict with the experimental result
of the branching ratio~\cite{pdg}: Br($D^+\to \pi^+\pi^0)= (1.24\pm 0.07) \times 10^{-3}$.
We conclude that $\eta_{(6,1)}$ is problematic to explain
the CDF $W+$dijets excess,
though the quantities can be adjusted by
choosing the possible couplings to right-handed quarks.

\subsubsection{The $({\bf3},{\bf3},-1/3)$ scalar}

In the case of $\eta_{(3,3)}$,
a similar analysis as in the previous section can be done
by supposing
$Y^{12}_{(3,3)}\neq0$ and
$Y_{(3,3)}^{23} = Y_{(3,3)}^{13}=0$ in the $Q^\prime = (V^\dagger U_L,D_L)$ basis.
In this case, $W$-$\eta$ box diagram
for the $K^0$-$\bar K^0$ mixing
vanishes
due to the color anti-symetricity,
\begin{equation}
\epsilon_{\alpha\beta\gamma}\epsilon^{\rho\eta\gamma}
= \delta_\alpha^\rho \delta_\beta^\eta
  - \delta_\alpha^\eta \delta_\beta^\rho,
\end{equation}
and only $\eta$-$\eta$ box diagram contributes. As a result we have:
\begin{eqnarray}
M_{12}^{K}(\eta)
 =
 \frac{f_K^2m_K }{48 \pi^2 M^2_W}
 \left(
 + \frac{3 \hat{B}_4 + \hat{B}_5}{4}
 \left(\frac{m_K}{m_s+m_d} \right)^2
 \frac{(Y^{12}_{(3,3)})^4}{32} E_5
\right).
\end{eqnarray}

For $Y_{(3,3)}^{12}=0.62$ and $m_{\eta} = 150$ GeV
which is chosen from the $W+$dijet excess,
the box contribution is the same size of the short distance SM contribution.
The coupling can also contribute to the strangeless charm decay width about 20\%.
While those quantities may be allowed within hadronic uncertainty, they nevertheless
push this scenario to the allowed boundary.

\subsubsection{The $({\bf3},{\bf1},-1/3)$ Scalar}

In the case of $\eta_{(3,1)}$ diquark,
the diquark coupling is a symmetric matrix,
\begin{equation}
\frac12 \overline{Q_L^c} Y_{(3,1)} \eta_{(3,1)} Q_L
= \overline{U_L^c} Y_{(3,1)} V \eta_{(3,1)} D_L.
\end{equation}
The $W$-$\eta$ box contribution also vanishes
due to the color anti-symmetricity.
The contribution to $K^0$-$\bar K^0$ mixing amplitude is
\begin{eqnarray}
M_{12}^{K}(\eta)
 =
 \frac{f_K^2m_K }{48 \pi^2 m^2_\eta}
 \left(
 + \frac{3 \hat{B}_4 + \hat{B}_5}{4}
 \left(\frac{m_K}{m_s+m_d} \right)^2
 \frac{X_{ij}}{8}
 F(z_i,z_j,1)
\right),
\end{eqnarray}
where
\begin{equation}
X_{ij} = (Y_{(3,1)}V)^*_{i1}(Y_{(3,1)}V)_{i2}
         (Y_{(3,1)}V)^*_{j1}(Y_{(3,1)}V)_{j2}.
\end{equation}
If we take $Y_{(3,1)} = y_{(3,1)} I$,
$y_{(3,1)} = 0.5$ and $m_\eta = 150$ GeV,
the $\eta$ contribution is twice as much as the short distance SM contribution.
However,
$Y_{(3,1)}$ is a symmetric matrix,
and thus, one can choose $(Y_{(3,1)}V)_{12}$ and
$(Y_{(3,1)}V)_{21}$ to be zero
to eliminate the flavor changing process.
(Under the choice, $(Y_{(3,1)}V)_{22} \simeq -(Y_{(3,1)}V)_{11}$.)
Therefore, the mixing amplitudes can be consistent with experiments.
There is no contribution to strangeless charm decay
in this choice.

We note that the color triplet bosons, $\eta_{(3,1)}$ and $\eta_{(3,3)}$,
can also have a leptoquark coupling $\overline{q_L^c}\ell \eta^*$
in general,
and it causes a severe problem of inducing too rapid nucleon decays.
One can avoid the rapid proton decays by introducing
a symmetry \cite{Dreiner:2006xw,Ajaib:2009fq},
allowing a milder baryon number violating process, such as neutron-antineutron
oscillations which can be tested at near future experiments \cite{Mohapatra:2009wp}.

We summarize the results in Table \ref{Table1} for the Yukawa couplings of the
colored scalars and FCNC constraints.
We conclude that there are scenarios which are consistent with FCNC data. Other ways of distinguishing these scenarios
should be studied. In the next section, we will study possible signatures at the RICH and LHC.

\begin{table}[tbp]
 \begin{tabular}{|c|c|c|c|c|c|c|}\hline
  & $\begin{array}{c}
      ({\bf8},{\bf2},1/2) \\
       {\rm with}\  Y_{8u}
     \end{array}$
  & $\begin{array}{c}
      ({\bf8},{\bf2},1/2) \\
       {\rm with}\  Y_{8d}
     \end{array}$ &
  $({\bf6},{\bf3},1/3)$ & $({\bf6},{\bf1},1/3)$ &
  $({\bf3},{\bf1},-1/3)$ & $({\bf3},{\bf3},-1/3)$ \\ \hline
  Flavor index
  & Arbitrary&Arbitrary& Symmetric & Anti-symmetric & Symmetric & Anti-symettric \\ \hline
  CDF $W+$dijet
  &$Y^{ij}_{8u} = 0.13\delta^{ij}$&$Y^{ij}_{8d}=0.19\delta^{ij}$& $Y^{11} = 0.32 $ & $Y^{12} = 1 $ & $Y^{11} = 0.5$ & $Y^{12} = 0.62$ \\ \hline
  FCNC &OK& OK &Fine tuning & Problematic & OK & Boundary
  \\ \hline
 \end{tabular}
 \caption{List of eligibility from the FCNC constraints
          of
          the couplings to explain the CDF $W$+dijets.}
 \label{Table1}
\end{table}

\section{Production of colored scalar at the RHIC and the
 LHC}\label{Sec:RHIC}

In this section, we study some implications for the
colored scalars which explain the CDF dijet excess at
the RHIC and the LHC.

Because the color sextet and triplet scalars can couple to di-quarks,
$pp$ colliders are suitable to search them from the resonance signal.
Since the mass of the scalar is not large, the $pp$ collider with low
center of mass energy has an advantage to avoid huge QCD backgrounds,
such as at the RHIC.

In hadroproduction of single heavy particle with the mass $m$, the mean
value of the energy fraction of partons inside proton is $\langle
x\rangle\sim\sqrt{\tau}$ where $\tau\equiv m^2/s$.
For the production of $\eta$ with $m_{\eta}=150$~GeV at the RHIC with
$\sqrt{s}=500$~GeV,  one has $\langle x\rangle\sim 0.3$, thus we expect
valence-valence quarks contribution brings large cross-section.

The diquark resonance signal can be observed as an excess in the
inclusive dijet events around $m_{jj}\sim m_{\eta}$.
We estimate the single diquark resonant production cross-section at the
RHIC.
The obtained cross-sections are listed in Table~\ref{Table2}.
The diquark-type scalars has a large cross-section of several tens to
hundreds pb.
On the other hand, for the color-octet scalar case, the cross-section is
only a few pb.

The main background comes from QCD processes which have broad
dijet invariant-mass distributions.
However, jets from the QCD processes have relatively large
pseudo-rapidity and small transverse momentum distributions, selection
cuts of, for example, $|\eta_{j}|<0.5$ and $p_{T,j}>50$~GeV can enhance
the signal to background ratio.

\begin{figure}[t]
\begin{center}
\includegraphics[width=.5\textwidth,clip]{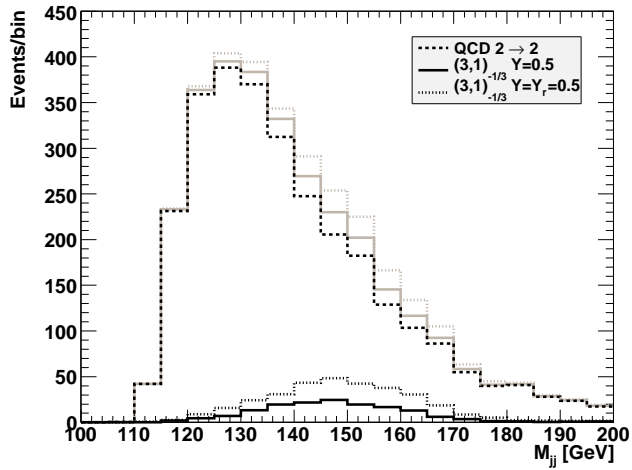}
\caption{Dijet invariant-mass distribution at the RHIC, for the exactly
 two-jet events with $|\eta_{j}|<0.5$ and $p_{T,j}>50$~GeV.
$L=10$~pb$^{-1}$ of the integrated luminosity is assumed.
Background event (dashed) is estimated by the 2$\to$2 QCD processes
 without $K$-factor correction.
Signal events in $\eta_{(3,1)}$ case are estimated without couplings to
 right-handed quarks (solid), and with couplings to right-handed quarks
 with $Y_r=0.5$ (dotted).
\label{RHIC}}
\end{center}
\end{figure}

For the $\eta_{(3,1)}$ case, it can give an excess to the
dijet invariant-mass distribution by roughly $S/N \sim 1/5$.
In Fig.~\ref{RHIC}, we plot the dijet invariant-mass distribution at the
RHIC with $\sqrt{s}=500$~GeV, after selecting the two-jet events with
the above pseudo-rapidity and transverse-momentum cuts for the jets.
The QCD background is estimated by QCD 2$\to$2 processes without
$K$-factor correction.
The number of the event is adjusted to the integrated luminosity of
$L=10$~pb$^{-1}$, which is already collected in 2009.
Since the accessible integrated luminosity at the RHIC is an order of
hundreds pb$^{-1}$, it should be possible to distinguish the excess from
the backgrounds even if one includes theoretical and experimental
uncertainties.\\
\begin{table}[tbp]
 \begin{tabular}{|c|c|c|c|c|c|c|}\hline
  & $\begin{array}{c}
      ({\bf8},{\bf2},1/2) \\
       {\rm with}\  Y_{8u}
     \end{array}$
  & $\begin{array}{c}
      ({\bf8},{\bf2},1/2) \\
       {\rm with}\  Y_{8d}
     \end{array}$ &
  $({\bf6},{\bf3},1/3)$ & $({\bf6},{\bf1},1/3)$ &
  $({\bf3},{\bf1},-1/3)$ & $({\bf3},{\bf3},-1/3)$ \\ \hline
  RHIC inclusive [pb]& 2.8 & 4.0 & 70 & 85 & 194 & 28 \\ \hline
  LHC $W+\eta $ [pb]& 8.8 & 15 & 75 & 70 & 42 & 50 \\ \hline
  LHC $Z+\eta $ [pb]& 0.8 & 1.5 & 3.0 & 23 & 13 & 8.6 \\ \hline
  Couple to $u_R d_R$ & Forbidden & Forbidden & Forbidden & Allowed &
  Allowed & Forbidden \\ \hline
 \end{tabular}
 \caption{List of the inclusive production cross sections at the RHIC and
          the $W$/$Z$ boson associate production at the LHC.
          At the RHIC, the diquark couplings to right-handed quarks
          can be also tested.}
 \label{Table2}
\end{table}

The triplet diquark $\eta_{(3,1)}$ can couple with right-handed quarks
by Yukawa-type interaction,
\begin{eqnarray}
- L=
 \overline{U^{c}_{R}}^{\alpha} Y_{r} \eta^{\beta}_{(3,1)}D_R^{\gamma}
 \epsilon_{\alpha\beta\gamma}+h.c.
\end{eqnarray}
The coupling $Y_r$ is generally independent from the couplings to the
left-handed quarks.
Although the $W+\eta$ production cross-section is unchanged by the
right-handed quark coupling, the single $\eta$ production
cross-section can be increased.
In Fig.~\ref{RHIC}, we also show how the cross-section would change by
introducing the coupling to right-handed quarks for $\eta_{(3,1)}$ case.
Assuming the same size coupling $Y^{11}_r=Y^{11}_{(3,1)}=0.5$ to the
first generation quarks, $q_{R}q_{R}$ scatterings give the same size
cross-section as the $q_{L}q_{L}$ scatterings, as easily expected.

Couplings to right-handed quarks are also possible for
$({\bf6},{\bf1},1/3)$ case, but forbidden in color-octet,
$({\bf6},{\bf3},1/3)$ and $({\bf3},{\bf3},-1/3)$ cases.
Note that $W+\eta$ production cross-section has no dependence on the
couplings to the right-handed quarks, but $Z+\eta$ production
cross-section has small dependence on the  couplings to the right-handed
quarks, because $Zq_Rq_R$ couplings are smaller than the $Zq_Lq_L$
couplings.

At the RHIC, using the polarization of the proton
beam~\cite{Bunce:2000uv}, it is possible to test the chiral structure of
the diquark couplings to quarks.
The partonic spin asymmetry, defined as
\begin{equation}
 \hat{a}=\frac
 {\hat{\sigma}_{LL} - \hat{\sigma}_{RR}}
 {\hat{\sigma}_{LL} + \hat{\sigma}_{RR}},
\end{equation}
where the subscripts describe the parton's helicity (chirality),
is found to be $\hat{a}=((Y^{11}_{(3,1)})^2 -
(Y^{11}_r)^2)/((Y^{11}_{(3,1)})^2 + (Y^{11}_r)^2)$ for the case we
consider.
Thus, it can probe the ratio of the left-handed coupling
$Y^{11}_{(3,1)}$ which is fixed by the CDF $W+$dijet excess, and the
right-handed coupling $Y_r$ which is unknown yet.
Using the knowledge of the polarized parton distribution functions of
quarks in valence distribution regions, it is possible to extract
the partonic spin asymmetry from the hadronic observables.
However, the detailed study is beyond the scope of this paper.

At the LHC, $W$+$\eta$ or $Z$+$\eta$ process followed by $\eta\to jj$
decay can be the signal again.
The production cross-sections at the LHC with
$\sqrt{s}=7$~TeV are also listed in Table~\ref{Table2}.
The expected major backgrounds are similar to those at the Tevatron;
$W/Z+$jets, $t\bar t$ and single-top production.
Detailed studies for the signal-to-background analysis at the LHC can be
found in Refs.~\cite{Eichten:2011xd,Harigaya:2011ww,Pappadopulo:2011rp},
for example.
The $W$+$\eta$ or $Z$+$\eta$ processes have large cross-sections as can
be seen from Table~\ref{Table2}, especially for the diquark-type models.
Following the study in Ref.~\cite{Pappadopulo:2011rp}, by taking into
account the QCD $W+$jets background, the $\ell\nu jj$ signal in the
$\eta_{(3,1)}$ case can be seen with the signal-to-background ratio of
$\sim0.12$ for the events with $120<M_{jj}<160$~[GeV].
Assuming the total detection efficiency to be $\sim0.05$, an expected
integrated luminosity for the $5\sigma$ discovery is
$\sim0.5$~[fb$^{-1}$] in this case.
For the color-octet scalar cases, the signal-to-background ratio is
estimated to be $\sim0.03$, therefore a better understanding of the
background events is needed to find the signal.

\section{Summary}\label{Sec:Sum}

We have studied the possibility of explaining the CDF $W+$dijet excess
by introducing colored scalar $\eta$ bosons.
Being colored scalars, through coupling to two quarks, they naturally decay
into dijet which provides one of the key feature of the $W+$dijet
excess.
There are several colored scalars,  $({\bf8}, {\bf2}, 1/2)$, $(\bar
 {\bf6}( {\bf3}) , {\bf3}({\bf1}), -1/3)$, $(\bar {\bf6}({\bf3}),
 {\bf1}, -4/3(2/3))$, which can have tree level
renormalizable Yukawa couplings with two quarks.
Not all of them can successfully explain the $W+$dijet excess.
Because the $W+$dijet excess requires a sizable coupling to the first
generation of quarks compared to the Higgs couplings to them,
the sizable couplings must also be consistent with other existing
experimental data.
We have analyzed FCNC constraints from meson-antimeson data.
We find that without forcing of the Yukawa couplings to be some special
texture forms most of the scalars, except the $({\bf 3}, {\bf 3},
-1/3)$, are in trouble with FCNC data.
We, however, find that the $({\bf 8}, {\bf 2}, 1/2)$ , $({\bf 6},{\bf
3}, 1/3)$ and $({\bf 3}, {\bf 1}, -1/3)$ can be made consistent with all
data.

While we confined our study to phenomenological implications of these
colored particle, we note that a concrete realization of their coupling
is harder to achieve.  Even one finds a flavor symmetry to forbid
certain entries of the Yukawa matrices, for example the off--diagonal
entries of $Y_{8u}$ for the octet, it is often the case that they are
induced at loop level. In this sense, all the scenarios discussed here
are to be considered as fine--tuned until a concrete realization is
achieved.

The candidate of the color triplet scalar is
an $SU(2)_L$ singlet,
and it also produces $Z+$dijet excess
at about 1/4 of the $ZZ+ZW$ process, which is not observed as a bump
around 150 GeV yet.
We also studied some predictions for the
diquark signal at $pp$ colliders, the RHIC and the LHC.
If the CDF excess is the diquark origin,
it may be confirmed at the early LHC study.
The RHIC experiment can help to distinguish the diquarks.

This work is partially supported by NSC, NCTS, SJTU 985 grant, and
Excellent Research Projects of
National Taiwan University (NTU-98R0526).
\\
\\

\noindent
{\bf Note Added}

After finishing this work, the CDF reported an updated
analysis~\cite{CDF2} using data collected through to November 2010
corresponding to an integrated luminosity of 7.3 fb$^{-1}$. Their
results are consistent with their early analysis~\cite{CDF} and
increased the significance to 4.1$\sigma$. Recently D0 collaboration
also reported their results of an analysis~\cite{Abazov} with an
integrated luminosity of 4.3 fb$^{-1}$. They did not find similar
$W+$diget excess. Although D0 was also looking at similar excess, the
methodology differs in some way which may be potentially important cause
for differences. We are not in a position to decide which one may be
correct which has to be settled among the experimental groups. We think
that a study of implications of the CDF results is still worthy. Our
results are not altered by the new CDF data.

\end{document}